\documentstyle[aps,prl,multicol]{revtex}
\input{psfig}
 \renewcommand{\[}{\begin{equation}} %
\renewcommand{\]}{\end{equation}} 
\newcommand{\eq}[1]{Eq.~(\ref{#1})} 

\newcommand{\llabel}[1]{\label{#1}}

\begin{document}
\title{Shot noise in long conductors}
\author{Yehuda Naveh}
\address{Department of Physics and Astronomy, State University 
of New York, Stony Brook, NY 11794-3800}
\date{\today }
\begin{multicols}{2}

\maketitle

{\bf Shot noise is the only fundamental type of current fluctuations
which exists in current-carrying conductors at low
temperatures.  As was recently stressed by Landauer\cite{Landauer 98},
it contains unique and important information about the correlations
which affect the electronic transport in the conductor.  
It is generally accepted that shot noise is a 'mesoscopic'
phenomenon in the sense that it vanishes when the length of the
conductor is much larger then the electron-phonon thermalization
length\cite{de Jong 96}. We show here that at least in one geometry
this result is valid only when the observation frequency $\omega$ is
strictly zero. At any finite frequency the shot noise, as measures in
the electrodes connecting the conductor, resumes its
value of the order of the 'full' shot noise $2eI$ ($I$ is the average
current in the conductor) provided that the sample is {\it long}
enough, $L > L_0 (\omega)$. Since even 'zero frequency' measurements
of shot noise are always done at RF, we claim here that
for sufficiently long samples (typically longer than a few
centimeters) such measurements should yield a large shot noise value.}

The basis for the calculations leading to the above conclusion is the
'drift-diffusion-Langevin' theory developed in Refs.~\cite{Naveh
97,Naveh 98} and valid at $\omega \ll 1/\tau, eV/\hbar$ with $\tau$
the elastic scattering time and $V$ the applied voltage. According to
this theory the noise spectral density as 
measured in the electrodes connecting the conductor (of length $L$
centered around $x=0$) can be presented as
\[
\label{noise}S_I(\omega)=\frac{2G}{L}\int_{-\frac L2}^{\frac L2
}|K(x;\omega )|^2{\cal C}(x)\,dx
\]
with $G$ the dc conductance and ${\cal C}(x)$ the correlator of local
fluctuations. This correlator depends only on the local distribution
function of electrons. For example, for an equilibrium
Fermi-Dirac distribution with temperature $T$ ${\cal C}(x) = 2T$, and
the noise assumes the 
Johnson-Nyquist value. For typical non-equilibrium distributions ${\cal
C}(x)$ becomes much larger than the equilibrium correlator, and at $T=0$
the noise 
given by \eq{noise} is the shot noise.

The response function $K(x; \omega)$, which is solely responsible for
the frequency dispersion of the noise, gives the current generated in
the electrodes by a fluctuating unit current source at $x$.  It is
dependent upon the specific geometry of the conductor and its
electrodynamic environment, but its integral over the sample length
always equals 1. We will study here a simple geometry for which an
analytical expression for $K(x; \omega)$ can be obtained, namely, a
thin and long conductor close to a ground plane. It is assumed that
both thickness of the conductor (in the direction perpendicular to the
ground plane) and its distance from the ground plane are much smaller
than $L$\cite{note coil}. Then the response function is given
by\cite{Naveh 98}
\[
\label{response}K(x;\omega )=\kappa \frac L2\frac{\cosh (\kappa
x)}{{\rm sinh}(\kappa L/2)}.
\]
Here $\kappa(\omega) =\sqrt{-i\omega /D^{\prime }}$ with $D^{\prime }=D+ G
L/C_0$ where $D$ is the diffusion coefficient and 
$C_0$ is the (dimensionless) linear capacitance between the conductor and
the ground plane. 

It is clear from \eq{response} that the only current fluctuations in
the conductor which are of importance in inducing noise in the
electrodes are those which are within a distance $\lambda_\omega =
1/\left| 
\kappa(\omega) \right|$ from the
conductor-electrode interfaces. Therefore, at high enough
frequencies, the measured noise is associated with the highly
non-equilibrium distribution of electrons near the edges of the
conductor, and not necessarily with the distribution at the
bulk of the sample  
(in long samples, the latter can be very close to an equilibrium
distribution, a fact which led to the common belief that the shot
noise should vanish in long samples). This simple argument means that
whenever $\lambda_\omega$ is smaller than some length scale  $l_S$
(which gives the spatial extent of non-equilibrium electrons in the
conductor -- see Fig.~1), the 
shot noise value should remain significant even with increasing $L$. 

\begin{figure}[tb]
\vspace{0.5cm}
\centerline{\hspace{0pt} \psfig{figure=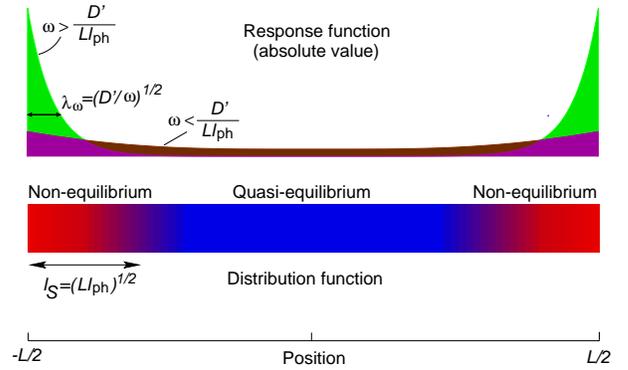,angle=-90,width=80mm}}
\vspace{0.5cm}
\centerline{\caption{\narrowtext
Schematic description of the relevant length scales in the problem:
The upper curve shows the response function for the two cases of
relatively low (brown) and high (green) frequencies. The lower panel
shows the spatial extent of non-equilibrium electrons near the edges
of the conductor.}}
\label{1scales}
\end{figure}

In order to give a quantitative description of the above effect, one
should solve the Boltzmann equation for the non-equilibrium
distribution $f$. To this end, we follow the prescription of
Ref.~\cite{Naveh 98a}, where $f$ was calculated under the assumption
of longitudinal acoustic phonon scattering. It was shown in that work
that the width of the layer in which the electron distribution is far
from equilibrium is $l_S \approx \sqrt{L l_{\rm
ph}}$, with  $l_{\rm ph}$ the inelastic scattering
length of an electron due to emission of a phonon of energy
$eV$. Therefore, one should expect 
large shot noise if 
$\lambda_\omega \approx \sqrt{D' / \omega} < \sqrt{L l_{\rm 
ph}}$, or $L > L_0(\omega)$ with
\[	\llabel{L0}
L_0(\omega) = {D' \over l_{\rm ph} \omega}.
\]

In what follows we would be interested in relatively long
samples and low frequencies. Therefore we assume here that the
electron-electron scattering length is much shorter than
 $L$ and $\lambda_\omega$. Having numerical results for $f$ (and
thus for ${\cal C}$) in this
situation\cite{Naveh 98a}, 
we can find the 
noise spectral density by combining Equations
(\ref{noise}) and (\ref{response}). 

Results for the noise spectral density $S_I(\omega)$ are presented in
Fig.~2 for a specific set of experimental parameters.
The upper curves in the figure show the total noise. The lower
curves show, on the same scale, the thermal noise. Since the latter is
smaller by an order of magnitude than the 
former, the upper curves actually depict the  shot noise. 

\begin{figure}[tb]
\vspace{0.3cm}
\centerline{\hspace{0cm} \psfig{figure=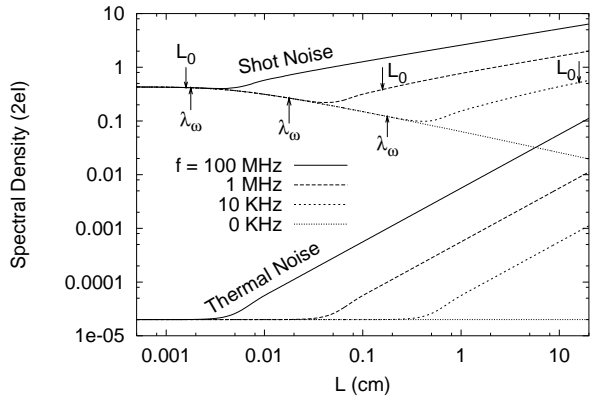,angle=-90,width=80mm}}
\vspace{0cm}
\centerline{\hspace{0cm} \caption{\narrowtext Noise spectral density as a function of sample's length. Lower
curves show the thermal noise and upper curves the full noise which is
dominated by the shot noise. Material parameters are $l_{\rm ph} =
10^{-3}$ cm and $D' = 1000$ cm$^2$/s. Temperature-to-voltage ratio
is $T/eV = 10^{-5}$. Arrows indicate the positions of $\lambda_\omega$
and $L_0$ for each of the depicted frequencies (at zero frequency
these lengths tend to infinity).}}
\label{2noise}
\end{figure}

The physical discussion presented above is fully supported by the
results shown in Fig.~2. One sees that at each of the three
frequencies depicted, the noise initially decreases with $L$ up to $L
\approx \lambda_\omega$, whereupon it increases, and reaches its
mesoscopic value again at $L \approx L_0(\omega)$. The initial
decrease of the noise with increasing $L$ is due to the electrons
being increasingly thermalized in the bulk of the sample, while the
subsequent increase is due to the widening of the non-equilibrium
surface layer as $\sqrt{L l_{\rm ph}}$, and therefore the increasing
distance from equilibrium of the noise-inducing electrons within the
layer of distance $\lambda_\omega$ from the interfaces (see Fig.~1).
As expected, at strictly zero frequency the noise reduces
monotonically to the thermal value at $L \rightarrow \infty$.

The unusual result of shot noise increasing with increasing sample
length is essentially due to a competition between two independent
physical processes: screening and equilibration. The importance of
screening in affecting shot noise was first discussed by Landauer in
qualitative terms\cite{Landauer 95,Landauer 96}, and was later studied
quantitatively in Refs.~\cite{Naveh 97,Naveh 98}. Its outcome effect
is summarized by \eq{response} and by the upper panel of
Fig.~1. Equilibration, on the other hand (lower panel of Fig.~1) is
responsible for the surface layers of non-equilibrium electrons. The
fact that the width $l_S$ of this layer grows with $L$ is readily
understood\cite{Naveh 98a}: since the electron-phonon relaxation time
decreases strongly with the energy of the emitted phonon, at large
$L$, when the electric field in the conductor is small, an electron
entering the sample from the electrode must diffuse elastically for a
long distance before it is able to emit a phonon.

Parameters chosen in obtaining Fig.~2 (except, possibly, the
temperature) correspond to typical
experimental situations (c.f. Ref.~\cite{Liefrink 94}). In fact, many 'zero
frequency' experiments are performed at an actual frequency of 10 KHz or
higher.  A different choice  of the effective
diffusion coefficient $D'$ would leave the results
unaltered only if accompanied by a similar change of the frequency
(in other words, the only dependence of $S_I$ on $\omega$ and on $D'$
is through 
$\omega/D'$). While $D = 1000$ cm$^2$/s is easy to achieve
experimentally, the electrostatic term in $D'$, $G L/ C_0 \approx D (t d
/ \lambda_0^2)$, dominates 
if the thickness $t$ of the conductor or its distance $d$ from
the ground plane are larger than the static screening length
$\lambda_0$. Thus, the most promising
geometry for the observation of the results depicted in 
Fig.~2 is within a thin layer, possibly a two-dimensional electron
gas. With thicker conductors, the macroscopic shot noise would be large
only at higher 
frequencies, or longer samples [see \eq{L0}].

In order to keep $L_0$ reasonably small $l_{\rm ph}$ must be large. To
maintain $l_{\rm ph} = 10^{-3}$ cm, $V$ cannot be larger than 100
mV. It is therefore likely that in an actual experiment $T/eV$ would
not be smaller than $10^{-3}$. Then, at large $L$ and $\omega$, the
thermal noise may be as large as the shot noise, and a subtraction
scheme should be deployed in order to extract the shot noise values
from the measurements.

The response function (\ref{response}), and thus the results presented
in this work, are not necessarily valid for geometries different from
the one studied here. The question of whether any specific geometry
exhibits shot noise when the conductor is long enough reduces to the
question whether finite-frequency fluctuations in the bulk of the
conductor are sufficiently screened as to not induce current in the
electrodes. Theoretically, a detailed answer to this question may
involve difficult solutions of the Poisson equation. However, in a
charged Fermi system finite-frequency currents are known to be
screened beyond some typical length scale $\lambda_\omega'$ which does
not depend on $L$\cite{Pines 66}. On the other hand, the
'hot-electron' length scale $l_S=\sqrt{L l_{\rm ph}}$ is independent
of the specific geometry. Therefore, it is argued that in sufficiently
long samples of an arbitrary geometry $l_S$ is larger than
$\lambda_\omega'$, so the only important sources of noise are from the
non-equilibrium regions near the electrodes. Following the physical
discussion below \eq{response}, this implies that the qualitative
features of the results presented here may be of a general nature.

I am indebted to D. V. Averin and K. K. Likharev for many
fruitful discussions. The work was supported in part by DOE's Grant
\#DE-FG02-95ER14575.

\end{multicols}
\end{document}